\documentclass[doublecol]{epl2} 
\usepackage{graphicx}
\usepackage{amsmath}
\usepackage{amsfonts}
\usepackage{amssymb}
\usepackage{color,colortbl}
\usepackage{multirow}

\newcommand{\rb}{\mathbf{r}}
\newcommand{\abs}[1]{\left| #1 \right|}
\newcommand{\avg}[1]{\langle #1 \rangle}
\newcommand{\nb}{\mathbf{n}}
\newcommand{\Nc}{\mathcal{N}}

\newcommand{\mub}{\mu_{\rb}}
\newcommand{\quotes}[1]{\textquotedblleft #1\textquotedblright}

\definecolor{Gray}{gray}{0.9}

\title{Universal subdiffusion of nonlinear waves in two dimensions with disorder}

\author{T.V.~Laptyeva \thanks{E-mail: \email{lapteva@pks.mpg.de}} \and J.D.~Bodyfelt \and S.~Flach}
\shortauthor{T.V.~Laptyeva \etal}

\institute{Max-Planck-Institut f\"{u}r Physik komplexer Systeme -- N\"{o}thnitzer Stra\ss e 38, D-01187 Dresden, Germany}
\pacs{05.45.-a}{Nonlinear dynamics and Chaos}
\pacs{05.60.Cd}{Classical transport}
\pacs{63.20.Pw}{Phonons in crystal lattices - Localized modes}

\abstract{We follow the dynamics of nonlinear waves in two-dimensional disordered lattices with tunable
nonlinearity. In the absence of nonlinear terms Anderson localization traps the packet in space.
For the nonlinear case a destruction of Anderson localization is found. The packet spreads subdiffusively, and its second moment
grows in time asymptotically as $t^\alpha$. We perform fine statistical averaging and test theoretical predictions 
for $\alpha$. Along with a precise confirmation of the predictions in [Chemical Physics \textbf{375}, 548 (2010)], 
we also find potentially long lasting intermediate deviations due to a growing number of surface resonances of the 
wave packet.}

\begin{document}

\maketitle

\section{Introduction}
Anderson localization (AL) - the halt of wave propagation in random potentials due to exponentially localized modes - was theoretically predicted over 50 years ago 
\cite{anderson_absence_1958} and in the past decades, since observed within a variety of experiments, including optics 
\cite{wiersma_localization_1997,cao_random_1999,chabanov_statistical_2000,stoerzer_observation_2006} and matter waves \cite{schulte_cold_2006,roati_anderson_2008}. 
These two are of strong interest, in that AL can be strongly altered by nonlinear Kerr effects in disordered photonic 
lattices \cite{pertsch_nonlinearity_2004,schwartz_transport_2007,lahini_anderson_2008}, or atomic Bose-Einstein condensate interactions in optical lattices 
\cite{shapiro_expansion_2007,skipetrov_anderson_2008,billy_direct_2008,sanchez-palencia_disordered_2010,modugno_anderson_2010}.

Research within nonlinear disordered media largely focuses on wave packet evolution in one-dimensional (1-d) systems. Asymptotic subdiffusive spreading is observed. 
An extended debate of the characterizing power exponents 
\cite{shepelyansky_delocalization_1993,molina_transport_1998,pikovsky_destruction_2008,veksler_spreading_2009,mulansky_spreading_2010,iomin_subdiffusion_2010,basko_weak_2011}
appears to be clarified by the theoretical predictions and their numerical verifications in 
\cite{flach_universal_2009,skokos_delocalization_2009, flach_spreading_2010,laptyeva_crossover_2010,bodyfelt_nonlinear_2011}
for 1-d cases.
Most studies focus on quartic nonlinearities which correspond to two-body interactions. Motivated greatly by experiment, e.g. in liquid crystal optics \cite{mihalache_nonlinear_1989,christian_helmholtz_2007} or at BEC-BCS crossovers in ultracold Fermi gases \cite{bloch_many-body_2008,yan_dark_2011}, one may also 
parametrize the nonlinearity exponent. This was done for the case of 1-d systems in \cite{mulansky_localization_2009,skokos_delocalization_2009} again with a confirmation 
of the theoretical prediction given in
\cite{flach_spreading_2010}. The innovation here is to extend to two-dimensional (2-d) disordered systems. For such lattices that are multidimensional, 
disordered, and have variable nonlinearity exponents, spreading behaviors were broadly conjectured within \cite{flach_spreading_2010}. The aim of this letter is to (dis)proof  
the conjectures for asymptotic spreading in 2-d lattices with
tunable nonlinearity. We will also investigate the case of small nonlinearity exponents at which the theory predicts
an anomaly in the number of wave packet surface resonances which should grow with ongoing spreading. We will test the robustness of the theoretical
predictions in this regime as well, where subdiffusion competes with fingering resonance instabilities.

\section{Generalized models}
The first model scrutinized is the generalized disordered nonlinear Schr\"odinger equation (gDNLS), which in a discrete lattice reads
\begin{equation}
\mathcal{H}_D=\sum_\rb  \left[ \epsilon_\rb \left| \psi_\rb \right|^2 + \frac{2 \beta \left| \psi_\rb \right|^{\sigma+2}}{\sigma+2} - \sum_{\nb \in \Nc} \left( \psi_\rb\psi_\nb^{\ast} + \psi_\rb^{\ast}\psi_\nb \right) \right]. 
\label{eq:HD}
\end{equation}
Here $\psi_\rb$ are complex variables, where $\rb=\left(x, y\right)$ denotes a 2-d square lattice vector of integer components with the nearest neighbor set of $\Nc \subseteq \left\lbrace \rb\pm (1,0),\rb\pm (0,1)\right\rbrace$. The disorder appears in on-site energies $\epsilon_\rb$, which are uncorrelated random values drawn uniformly from an interval $[-W/2,W/2]$ parameterizing by the disorder strength $W$. The nonlinearity of strength $\beta$ is generalized to a power $\sigma > 0$. This is best seen in the equations of motion, derived from $\dot{\psi}_\rb = \partial \mathcal{H}_D / \partial(i\psi_\rb^{\ast})$ as
\begin{equation}
i \dot{\psi}_\rb = \epsilon_\rb \psi_\rb + \beta \left|\psi_\rb \right|^\sigma \psi_\rb - \sum_{\nb \in \Nc} \psi_\nb. \label{eq:HDEOM}
\end{equation}
The above set of dynamic equations conserves the total energy $\mathcal{H}_D$, as well as the total norm $S=\sum_\rb \left|\psi_\rb \right|^2$. The 1-d version of the gDNLS has been extensively studied: for $\sigma=2$ it relates to recent experimental photonics \cite{lahini_anderson_2008} and has been investigated numerically \cite{flach_universal_2009,skokos_delocalization_2009,flach_spreading_2010,laptyeva_crossover_2010,bodyfelt_nonlinear_2011}. For a few integer values of $\sigma$ 1-d simulations were presented in \cite{mulansky_localization_2009}. Simulations with non-integer $\sigma$ were also performed \cite{veksler_spreading_2009} on short time scales without focus on asymptotic spreading.

The second model considered is the generalized 2-d Klein-Gordon (gKG) lattice, governed by the Hamiltonian
\begin{equation}
\mathcal{H}_K=\sum_\rb \left[ \frac{p_\rb^2}{2} + \frac{ {\tilde \epsilon}_\rb u_\rb^2}{2} + \frac{ \abs{u_\rb}^{\sigma+2} }{\sigma+2}  + \frac{1}{2W}\sum_{\nb \in \Nc} \left( u_\nb-u_\rb \right)^2 \right], 
\label{eq:HKG}
\end{equation}
where $u_\rb$ and $p_\rb$ respectively are generalized coordinates and momenta on the lattice site $\rb$, with an energy density of $\mathcal{E}_\rb$. The terms ${\tilde \epsilon}_\rb$ are uncorrelated random values drawn uniformly from an interval $[1/2,3/2]$. From $\ddot{u}_\rb=-\partial\mathcal{H}_K/\partial u_\rb$, the equations of motion read 
\begin{equation}
\ddot{u}_\rb=-{\tilde \epsilon}_\rb u_\rb - \left|u_\rb\right|^\sigma u_\rb + \frac{1}{W}\sum_{\nb \in \Nc}\left( u_\nb - 4 u_\rb \right).
 \label{eq:HKGEOM}
\end{equation}
This set of dynamic equations conserve only the total energy $\mathcal{H}_K = \sum_\rb \mathcal{E}_\rb$. The 1-d version of the gKG has also been extensively studied, as it can be considered a model for dynamics of anharmonic optical lattice vibrations in molecular crystals \cite{ovchinnicov_from_2003}. The quartic 1-d case was heavily used in numerical investigations \cite{flach_universal_2009,skokos_delocalization_2009,flach_spreading_2010,laptyeva_crossover_2010,bodyfelt_nonlinear_2011}, and different values of $\sigma$ have also been addressed \cite{skokos_spreading_2010}. 
 
Several works \cite{kivshar_modulation_1992, johansson_statistical_2004, johansson_discrete_2006} suggest an equivalence between the two models. Note, that in KG there is no nonlinear parameter $\beta$. Rather the scalar value $\mathcal{H}_K$ acts as the nonlinearity control: by writing $ u_\rb$ as a plane wave in Eq.(\ref{eq:HKGEOM}) and applying slow modulation/rotating wave approximations, Eq.(\ref{eq:HDEOM}) can be recovered under an approximate condition. For quartic nonlinearity, this condition is $\beta S \approx 3W\mathcal{H}_K$. It connects the KG initial parameters $\mathcal{H}_K$ and $W$ to the total initial norm $S$ and nonlinear parameter $\beta$ of the corresponding quartic DNLS model. This condition can furthermore be generalized to any power $\sigma$
\begin{equation}
\beta \sum_\rb \left| \psi_\rb \right|^\sigma \approx a_\sigma W \sum_\rb \mathcal{E}_\rb^{\sigma/2}, \quad a_\sigma \equiv \frac{8(\sigma+1)\Gamma(\sigma)}{\sigma(\sigma+2)\Gamma^2(\sigma/2)}.
\label{eq:map}
\end{equation}
In this derivation, the absolute value in Eqs.(\ref{eq:HDEOM},\ref{eq:HKGEOM}) is ignored, since its inclusion was found only to yield minor higher order corrections to $a_\sigma$. A similar result for the nonlinear shifts in energy was seen in \cite{skokos_spreading_2010}. 

Neglecting nonlinear terms both Eqs.(\ref{eq:HDEOM},\ref{eq:HKGEOM}) reduce to an eigenvalue problem, giving a set of exponentially localized eigenstates (denoted as normal modes, NM) with frequencies $\lambda_\rb$ in a spectrum of width $\Delta_D=8+W$ in the case of gDNLS. Linear reduction for the gKG is similar, but with squared frequencies $\omega_\rb^2$ in a spectrum of width $\Delta_K=\Delta_D/W=1+8/W$. We will focus mainly on analytics of the gDNLS, 
since it is straightforward to adapt results for the gKG using Eqs.(\ref{eq:map}). 
\section{Expected spreading regimes}
Consider the time-dependent normalized norm density distribution, $z_\rb \equiv \left| \psi_\rb \right|^2 / S$. The gKG counterpart is the normalized energy density distribution, $z_\rb \equiv \mathcal{E}_\rb / \mathcal{H}_K$. Distributions are analyzed by means of the second moment, $m_2=\sum_\rb \left| \rb-\mub \right|^2 z_\rb$, where the density center is $\mub=\sum_\rb \rb z_\rb$. The second moment quantifies the squared width of the packet, hence, its spreading. The participation number, $P=1/\sum_\rb z_\rb^2$, measures the number of effectively excited sites. Lastly, the packet sparseness is measured by the compactness index \cite{flach_universal_2009}, which for 2-d models is $\zeta = P/m_2$. How then do these three measures behave for different parameters? 

In the linear case, the participation number approximates the spatial extension of a NM, with an average measure over modes being the localization volume $V$. The dependence of $V$ on $W$ is shown in the inset of Fig.\ref{fig:parspace}, where squares [diamonds] are for the linear version of Eq.(\ref{eq:HD}) [Eq.(\ref{eq:HKG})], the gray cloud is the overall standard deviation, and the solid line is a best fit of $V\sim W^\gamma$ through all points. Similar curves also appear in \cite{schreiber_localization_1992, zharekeshev_crossover_1996}. The average frequency spacing of NMs within the localization volume is $d=\Delta_D/V$. The two linear frequency scales $d$ and $\Delta_D$ are expected to contribute to the details of packet spreading. Nonlinearity also introduces an additional frequency scale - the nonlinear shift of a single oscillator, proportional to $\beta \rho^{\sigma/2}$ for the gDNLS, where $\rho$ is the average norm density of a packet. From these three unique frequency scales, dynamical regimes of packet spreading were presented in 1-d quartic systems \cite{flach_universal_2009,skokos_delocalization_2009,flach_spreading_2010,laptyeva_crossover_2010,bodyfelt_nonlinear_2011} for a variety of different initial parameters. Both the initial norm/energy density of a packet and its typical size were suggested \cite{flach_spreading_2010} as the major control parameters for the dynamics at given $W$ and $\sigma$. Under strong enough nonlinearity, a fraction of a wave packet (or even the whole packet) exhibits self-trapping \cite{kopidakis_absence_2008,laptyeva_crossover_2010,bodyfelt_nonlinear_2011}. For weaker nonlinearity (such that self-trapping is avoided), two possible dynamical outcomes are predicted. The packet spreads in an intermediate regime of strong chaos with subsequent dynamical crossover into an asymptotic regime of weak chaos, or spreading starts directly in the weak chaos regime. 

A straightforward generalization of these expected regimes of packet spreading, with an initial packet norm density $\rho$ and size $L < V$, was proposed \cite{flach_spreading_2010} as
\begin{eqnarray}
\beta \rho^{\sigma/2}(L/V)^{\sigma/2} < d &\quad & \mbox{weak chaos}, \nonumber \\ 
\beta \rho^{\sigma/2}(L/V)^{\sigma/2} > d &\quad & \mbox{strong chaos}, \label{eq:regimes} \\
\beta \rho^{\sigma/2} > \Delta_D &\quad& \mbox{self-trapping}. \nonumber
\end{eqnarray}

The spreading mechanism is thought to be an incoherent energy transfer between NMs inside the packet to nearby exterior NMs \cite{flach_universal_2009,flach_spreading_2010,krimer_statistics_2010}. The derivation of possible dynamical outcomes is based on the resonance probability $\mathcal{P}$ of this transfer; this in turn depends on the norm/energy density of the packet as $\mathcal{P}(\beta \rho^{\sigma/2}) \approx 1-\exp\left(-\beta \rho^{\sigma/2} d^{-1}\right)$. In the regime of weak chaos, only a small fraction of interior modes resonantly interact and $\mathcal{P} \approx \beta \rho^{\sigma/2} d^{-1}$. In the regime of strong chaos, nearly all modes interact, i.e. $\mathcal{P} \approx 1$. The diffusion rate $D$ is conjectured to be $D \sim \beta^2 \rho^\sigma (\mathcal{P}(\beta \rho^{\sigma/2}))^2$ which together with $m_2 \sim \rho^{-1}$, leads to power laws for spreading in 2-d
\begin{equation}
m_2, P \sim t^\alpha, \quad \alpha = \begin{cases}
          \frac{1}{1+2\sigma} & \mbox{weak chaos}, \\
	  \frac{1}{1+\sigma}  & \mbox{strong chaos}. 
         \end{cases} 
         \label{eq:powers}
\end{equation}
Note, since the packet norm/energy density decreases in time and eventually the condition for strong chaos will be no longer satisfied -- spreading will cross into the regime of weak chaos. Nevertheless, the duration of strong chaos regime can be greatly prolonged (over multiple orders of magnitude), so much that the crossover occurs at infeasible computation times.

In the regime of weak chaos, the number of resonances in the packet volume $N_{RV}$ and on its surface $N_{RS}$ are estimated \cite{flach_spreading_2010} as
\begin{equation}
N_{RV} \sim \beta \rho^{\sigma/2-1}, \quad N_{RS} \sim \beta \rho^{(\sigma-1)/2}. 
\label{eq:volsurf}
\end{equation}
According to the above equation, for the 2-d case there are critical values of nonlinearity power: the number of volume resonances will grow for any $\sigma < 2$, likewise $\sigma < 1$ for surface resonances. We therefore expect these critical values may manifest unusual effects in the course of packet spreading (we shall return to this point in the last section).
\section{Numerical Simulations}
Our following numerics present only gKG results, for two reasons. Firstly, the presence of a corrector scheme for gKG (e.g. Appendix of \cite{skokos_delocalization_2009}) allows two magnitude orders greater in integration, at the same conservations and integration speeds. Secondly, the gKG requires only a single conservation. Lastly, all prior simulations of both models \cite{flach_universal_2009,skokos_delocalization_2009,flach_spreading_2010,laptyeva_crossover_2010,bodyfelt_nonlinear_2011} show similar qualitative results in a wide range of energy and disorder.

\begin{figure}[htb]
\includegraphics[width=0.9\columnwidth,keepaspectratio,clip]{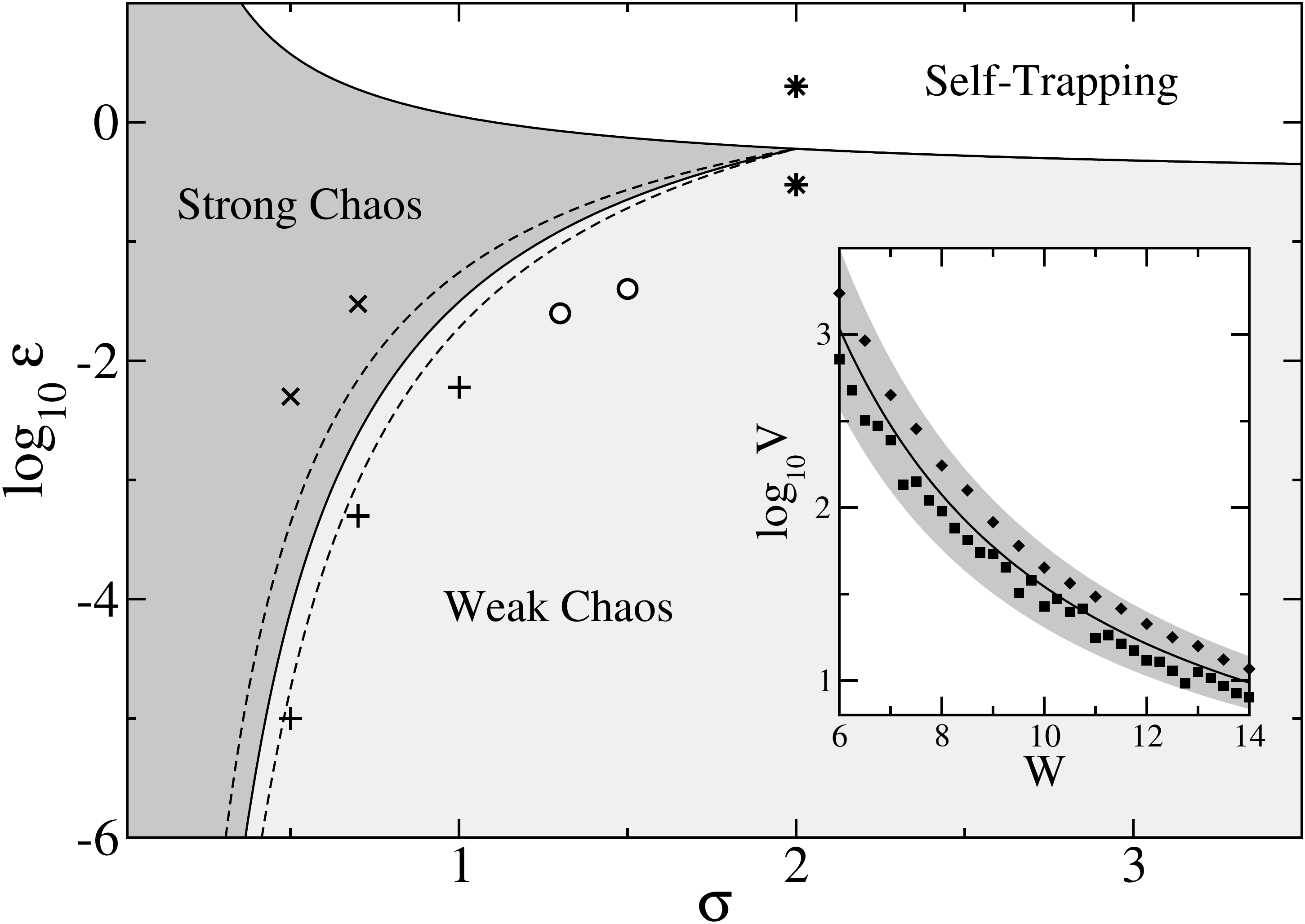}
\caption{\textit{Main:} parameter space of the nonlinearity power $\sigma$ and the
energy $\mathcal{E}$ of a single-site excitation ($L=1$). Dashed lines show the variation of the boundary obtained from variation of $V$ (see inset). For our numerics, the different symbols correspond to points $(\sigma,\mathcal{E})$ used. The various behaviors discussed in further sections of the text: \quotes{$\ast$} for $(2,0.3),(2,2.0)$,  \quotes{$+$} for $(0.5,0.00001),(0.7,0.0005),(1.0,0.006)$, \quotes{$\circ$} for $(1.3,0.025),(1.5,0.04)$, and \quotes{$\times$} for $(0.5,0.005),(0.7,0.03)$. \textit{Inset:} the dependence of localization volume $V$ on $W$. Squares [diamonds] are for the linear version of Eq.(\ref{eq:HD}) [Eq.(\ref{eq:HKG})]. The gray region denotes an overall standard deviation. The solid line is a best fit to $V\sim W^\gamma$.}
\label{fig:parspace}
\end{figure}

To test the analytical prediction of Eq.(\ref{eq:powers}), we first set the localization volume to $V \sim 34$ for $W=10$. This will stay the same for all presented numerics. For an initial single-site excitation ($L=1$) of energy $\mathcal{E}$, regime boundaries from Eq.(\ref{eq:regimes}) can be easily mapped into a $\mathcal{E}(\sigma)$ form using Eq.(\ref{eq:map}). This effectively gives a parameter space, shown in the main of Fig.\ref{fig:parspace}. The two dashed lines show the boundary deviation, due to variations in $V$ (see inset). 

We then select energy density $\mathcal{E}$ so that our initial state is set as $p_\rb = \sqrt{2\mathcal{E}}$ for $\rb=(N/2,N/2)$ and zero elsewhere, $u_\rb=0$ everywhere. This state is numerically integrated by Eq.(\ref{eq:HKGEOM}) using SABA-class symplectic integrators \cite{laskar_high_2001} under a time-step of $10^{-1}$ to maximums of $10^{6-8}$, all the while maintaining a relative energy conservation up to $10^{-2}$. For each set of parameters, we calculate the three measures $m_2, P, \zeta$. The typical lattice size is $200 \times 200$ sites. We then average over $400$ disorder realizations. For each parameter set, we determine the spreading power-law exponent as local derivative $\alpha(t) \equiv {\rm d} \left\langle \log_{10} m_2 \right\rangle / {\rm d} \log_{10}t$ (see, e.g. \cite{laptyeva_crossover_2010}).
\subsection{Subdiffusion in 2-d lattices with $\sigma=2$: weak chaos and self-trapping}
\begin{figure}[hbt]
\includegraphics[width=0.9\columnwidth,keepaspectratio,clip]{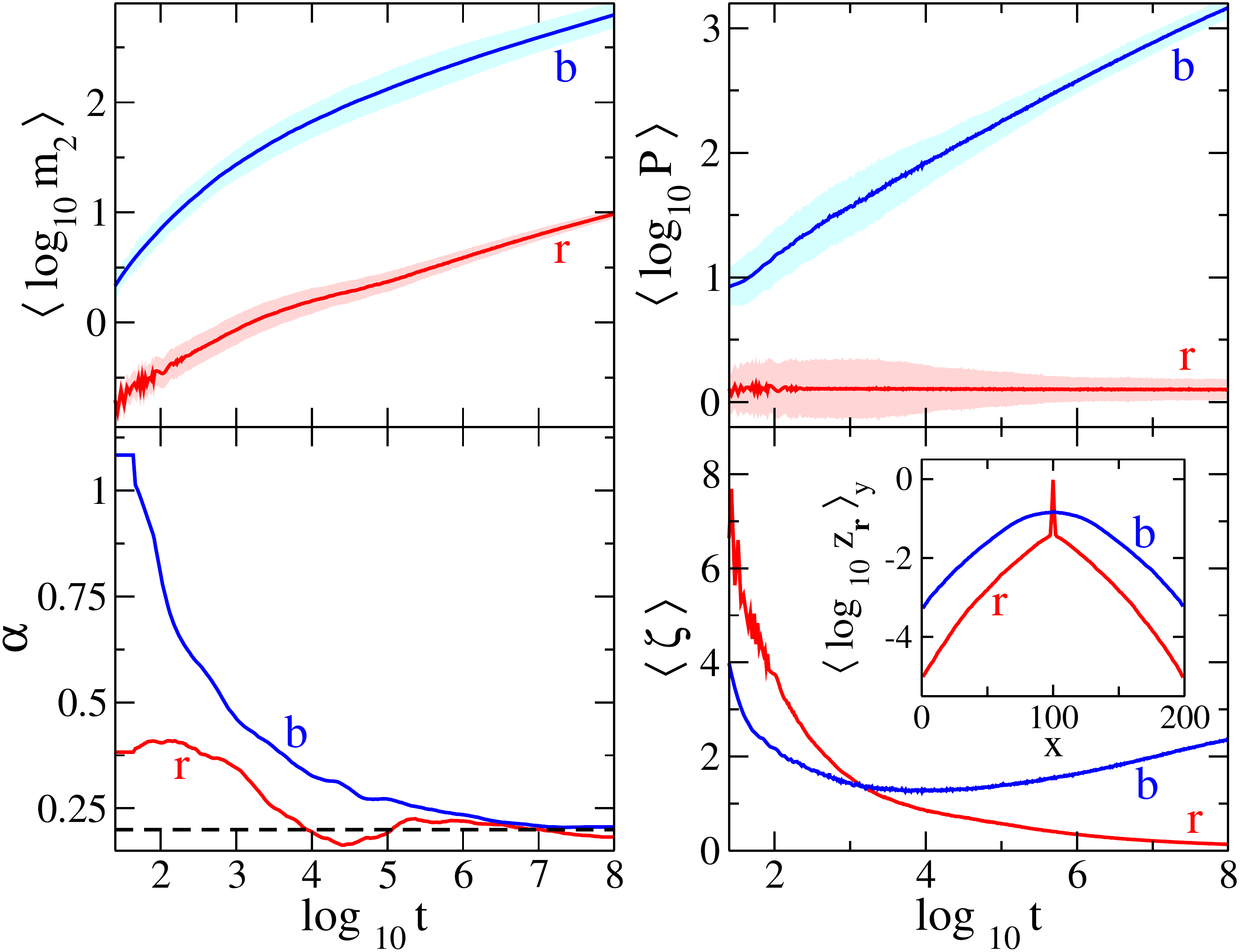}
\caption{(Color online.) Numerics for \quotes{$\ast$} in Fig.\ref{fig:parspace}. The parameters $(\sigma,\mathcal{E})=(2, 0.3), (2, 2.0)$ correspond to the weak chaos ((b)lue) and self-trapping ((r)ed). 
\textit{Left column}: second moment (upper) and its power-law exponent $\alpha$ (lower). The dashed line is our theoretical expectation for weak chaos $\alpha=1/(1+2\sigma)=0.20$.
\textit{Right column}: participation number (upper) and compactness index (lower). In both columns of the upper row the lighter clouds correspond to a standard deviation.
\textit{Inset:} Normalized density distributions at $t=10^8$, averaged first over realization and then over the $y$-coordinate. Self-trapping is clearly seen.}
\label{fig:sig_eq2}
\end{figure}

In this subsection, the numerics for quadratic nonlinearity (\quotes{$\ast$} in Fig.\ref{fig:parspace}) are discussed. With single-site excitations, the strong chaos regime is unreachable for $\sigma \geq 2$. In Fig.\ref{fig:sig_eq2}, the weak chaos regime ($\mathcal{E}=0.3$, lower \quotes{$\ast$} in Fig.\ref{fig:parspace}) displays spreading, with $m_2$ and $P$ growing (blue curves) to reach an asymptotic spreading following the predicted power law of Eq.(\ref{eq:powers}), seen by the saturation $\alpha\simeq 0.2055$ and verifying our theoretical prediction of $1/5$ for weak chaos. This ought be compared to an earlier work \cite{garcia-mata_delocalization_2009}, in which the asymptotic law was hypothesized as $1/4$, but only via integration to $10^6$. Indeed, at this time, we also approximately obtain the same value ($\alpha\simeq 0.234$); however upon integrating further, our expectation of $\alpha=1/5$ is revealed by the saturation. 
In this regime, the asymptotic compactness index is $\zeta\simeq 2.36$ (seen in the blue curve of Fig.\ref{fig:sig_eq2}), meaning that the packet spreads, yet remains largely thermalized ($\zeta \approx 3$). A trend of $\zeta \rightarrow 0$ indicates either very sparse packets or partial self-trapping \cite{laptyeva_crossover_2010}. Large values of $\mathcal{E}$ satisfy the self-trapping regime. This behavior can be seen in the red curves of Fig.\ref{fig:sig_eq2} ($\mathcal{E}=2.0$, upper \quotes{$\ast$} in Fig.\ref{fig:parspace}). A large portion of energy remains trapped on the initially excited site, while a much smaller portion subdiffuses. We thusly observe an increase of $m_2$ (smaller portion subdiffusing), while $P$ remains largely unaffected (large self-trapped portion). The compactness index approaches zero -- a very good indication of self-trapping. Lastly, in the inset of Fig.\ref{fig:sig_eq2}  we show normalized density distributions (averaged first over realization, then over the $y$-coordinate) for both regimes. The self-trapping regime reveals a characteristically trapped portion in the center. 
\subsection{Subdiffusion in 2-d lattices with $1<\sigma<2$: weak chaos}
\begin{figure}[hbt]
\center
\includegraphics[width=0.9\columnwidth,keepaspectratio,clip]{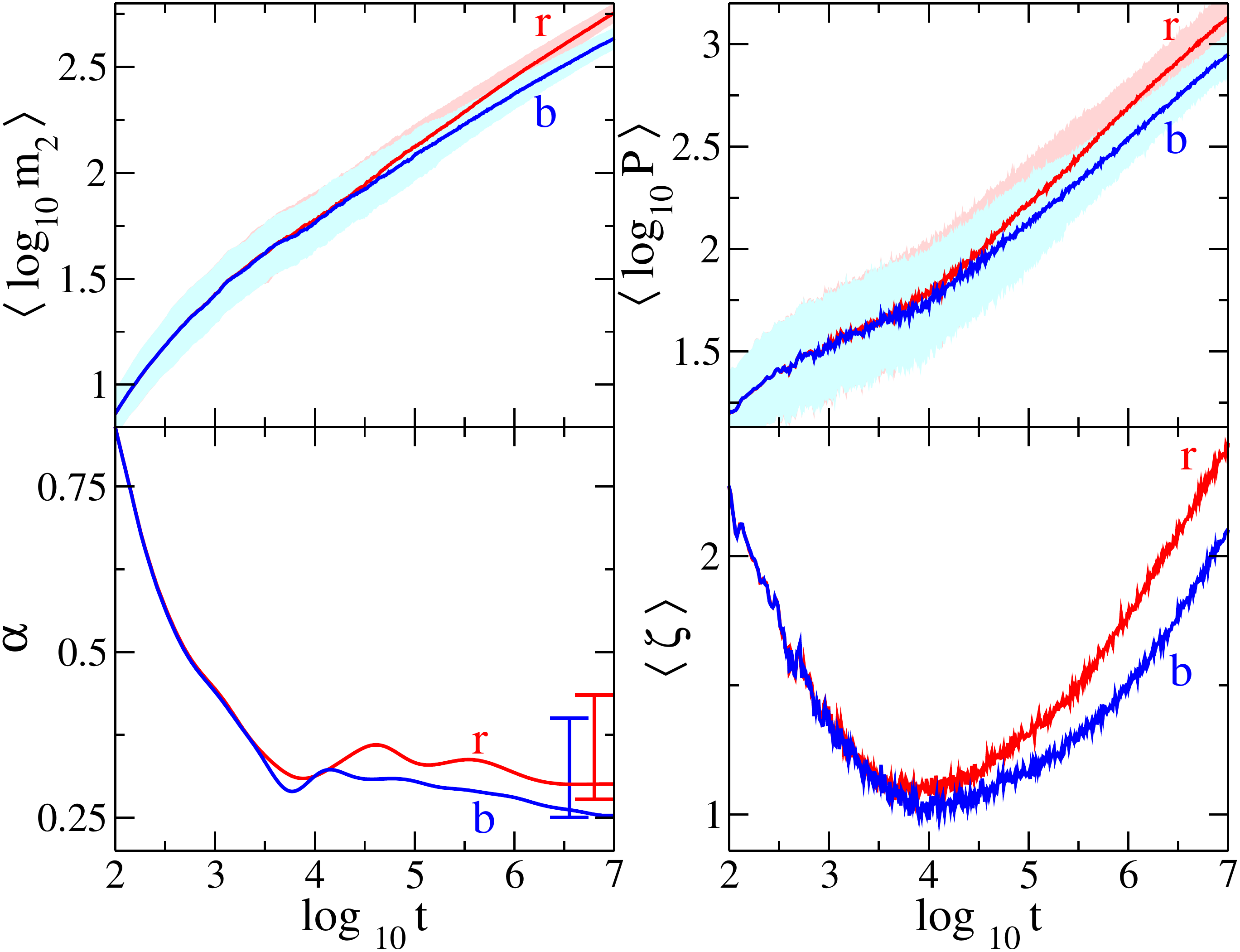}
\caption{(Color online.) Numerics for \quotes{$\circ$} in Fig.\ref{fig:parspace}. The parameters $(\sigma,\mathcal{E})=(1.3,0.025),(1.5,0.04)$ are colored respectively as (r)ed and (b)lue.    
\textit{Left column}: the second moment (upper) and its power-law exponent $\alpha$ (lower). The I-bar bounds denote the theoretical expectations from Eq.(\ref{eq:powers}) for weak chaos (lower bound) and strong chaos (upper bound).  
\textit{Right column}: participation number (upper) and compactness index (lower). In both columns of the upper row the lighter clouds correspond to a standard deviation.}
\label{fig:sig_gt1_lt2}
\end{figure}

Numerical findings for \quotes{$\circ$} in Fig.\ref{fig:parspace} with parameters $(\sigma,\mathcal{E})=(1.3,0.025), (1.5,0.04)$ are shown in Fig.\ref{fig:sig_gt1_lt2}. Particularly in the lower left panel, the exponent $\alpha$ is shown to also asymptotically saturate. The two \quotes{I-bars} do not show error per se, rather their lower/upper bounds dictate the weak/strong chaos expectations for $\alpha$ from Eq.(\ref{eq:powers}). 
The asymptotic saturations are approximately $\alpha\simeq 0.300$ for $\sigma=1.3$ and $\alpha\simeq 0.257$ for $\sigma=1.5$, which is quite close to their respective weak chaos expectations of $0.278$ and $0.250$. Additionally, compactness values of $\zeta\simeq 2.44,2.10$ at $t=10^7$ remain fairly thermalized. Therefore, these two points fully follow our expected theory of weak chaos spreading. 
\subsection{Subdiffusion in 2-d lattices with $\sigma\le 1$: strong chaos}
Moving to the left in Fig.\ref{fig:parspace}, we cross the theoretical division between strong and weak chaos (two representative points are shown by \quotes{$\times$}). Again, we perform our numerics for these two points. Note, that red curves are for $(\sigma,\mathcal{E})=(0.5,0.005)$ and blue curves are for $(\sigma,\mathcal{E})=(0.7,0.03)$. Similarly, the I-bar bounds give weak chaos (upper) and strong chaos (lower) expectations for the respective values of $\sigma$. The asymptotic saturations for these two representative points are approximately $\alpha\simeq 0.669$ for $\sigma=0.5$ and $\alpha\simeq 0.571$ for $\sigma=0.7$, which is quite close to their respective strong chaos expectations of $0.667$ and $0.589$.
Additionally, the compactness index fluctuates due to a slight asymptotic slope change in the participation numbers, nevertheless, it remains nearly thermalized about $1.7$ for $t=10^6$. Therefore, these two points follow our expected theory of strong chaos spreading. 
\begin{figure}[htb]
\center
\includegraphics[width=0.9\columnwidth,keepaspectratio,clip]{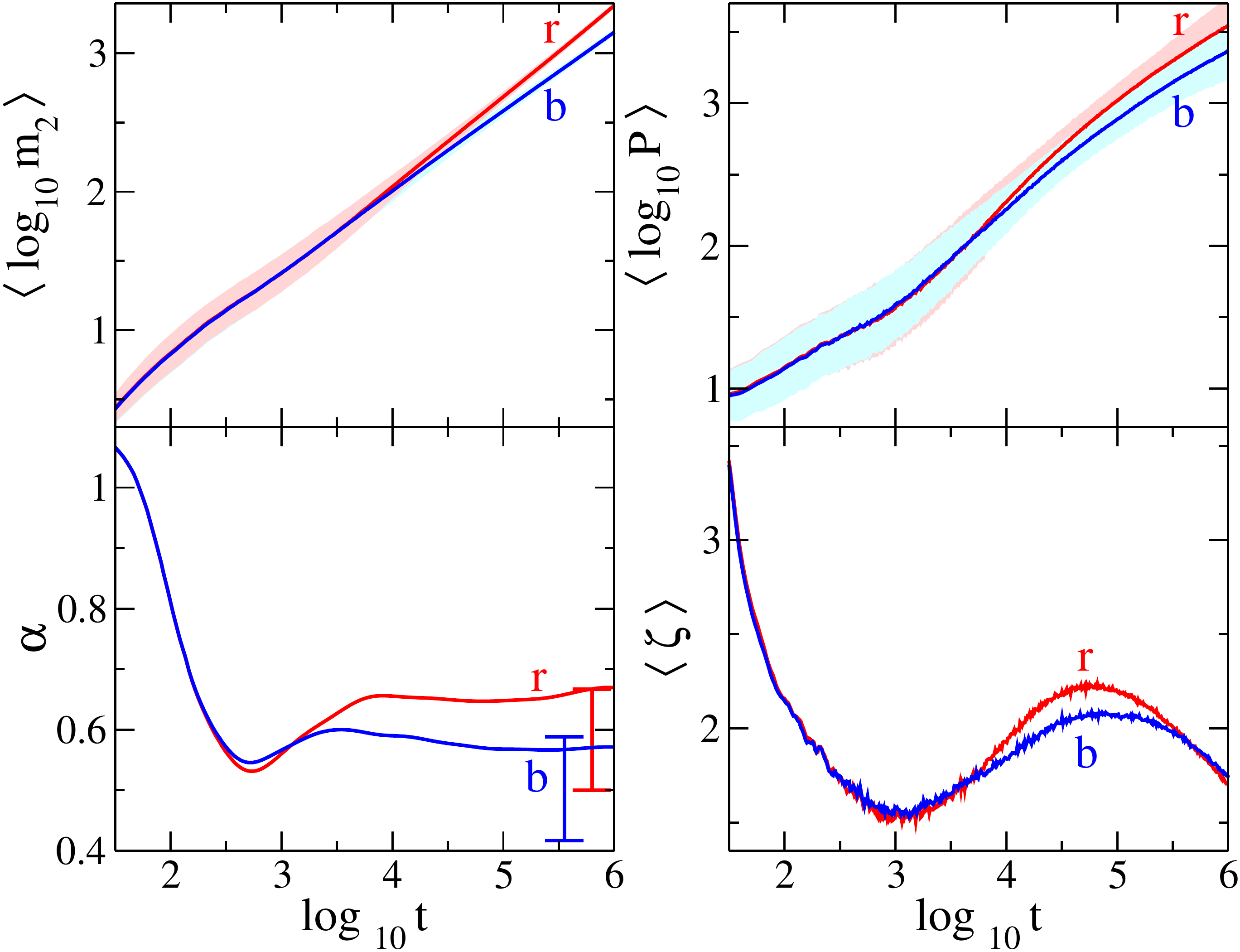}
\caption{(Color online.) Numerics for \quotes{$\times$} in Fig.\ref{fig:parspace}. The parameters $(\sigma,\mathcal{E})=(0.5,0.005),(0.7,0.03)$ are colored respectively as (r)ed and (b)lue.    
\textit{Left column}: the second moment (upper) and its power-law exponent $\alpha$ (lower). The I-bars denote the theoretical expectations from Eq.(\ref{eq:powers}) for weak chaos (lower bound) and strong chaos (upper bound).  
\textit{Right column}: participation number (upper) and compactness index (lower). In both columns of the upper row, the lighter clouds correspond to a standard deviation.}
\label{fig:sig_le1_sc}
\end{figure}
\subsection{Subdiffusion in 2-d lattices with $\sigma\le 1$: intermediate behaviors}
\begin{figure}[thb]
\center
\includegraphics[width=0.9\columnwidth,keepaspectratio,clip]{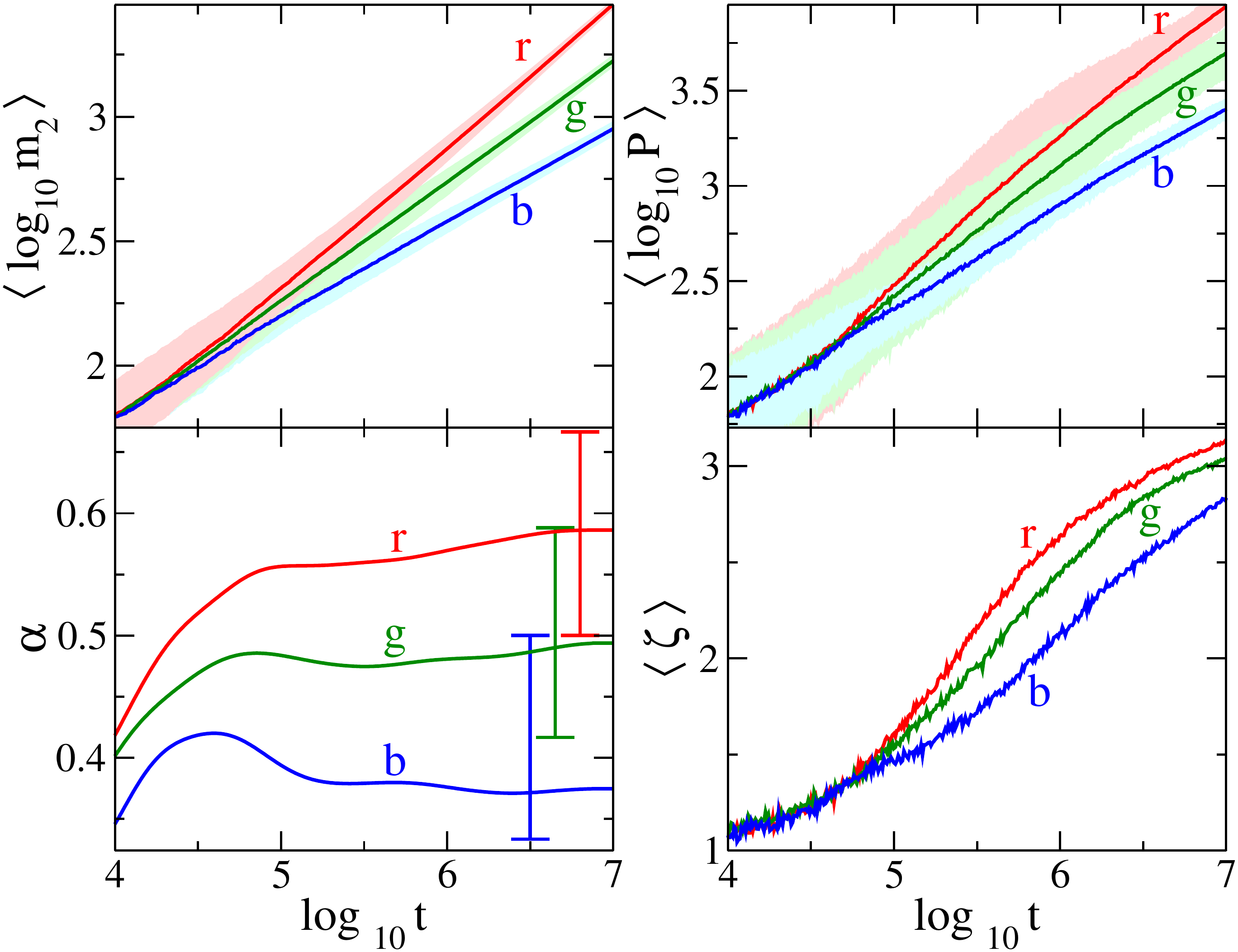}
\caption{(Color online.) Numerics for \quotes{$+$} in Fig.\ref{fig:parspace}. The parameters $(\sigma,\mathcal{E})=(0.5,0.00001),(0.7,0.0005),(1.0,0.006)$ are colored respectively as (r)ed, (g)reen, and (b)lue. 
\textit{Left column}: the second moment (upper) and its power-law exponent $\alpha$ (lower). Similarly, I-bar bounds denote the theoretical expectations from Eq.(\ref{eq:powers}) for weak chaos (lower bound) and strong chaos (upper bound). \textit{Right column}: participation number(upper) and compactness index (lower). In both columns of the upper row, the lighter clouds correspond to a standard deviation.}
\label{fig:sig_le1}
\end{figure}

Moving down in Fig.\ref{fig:parspace}, we cross back to the theoretical weak chaos regime (three representative points are given by \quotes{$+$}). Performed numerics are shown in Fig.\ref{fig:sig_le1}, where red curves are for $(\sigma,\mathcal{E})=(0.5,0.00001)$, green curves are for $(\sigma,\mathcal{E})=(0.7,0.0005)$, and blue curves are for $(\sigma,\mathcal{E})=(1.0,0.006)$. These points exhibit behavior novel from that was seen in the previous sections. Namely, in the lower left panel the exponent power $\alpha$ reaches asymptotic values of $0.586,0.494,0.375$ respectively for red, green, and blue curves. That is a clear tendency toward saturations resting in neither regime. 
Rather, these values lay between the two bounds from the expected regimes. Similar behavior was also hinted about $\sigma=1.0$ for 1-d, particularly Fig.5 of \cite{skokos_spreading_2010}. Returning to our argument of resonance probability, this suggests both the weak/strong limits are invalid - the value of $\mathcal{P}$ must be explicitly found. More than just a few modes contribute, but certainly not enough to yield strong chaos regime. 
\subsection{Dimensional Analysis}
According to Eq.(\ref{eq:volsurf}), the concept of resonance probability may be viewed in the light of competition between surface growth versus volume growth. Surface resonances more easily lead to density leakage into modes exterior to the packet. This process in turn increases the packet's perimeter, therefore yielding more surface resonances. Packets may thusly develop finger structures or fragment, perhaps leading to a fractal-like structure. 

\begin{figure*}[bht]
\center
\includegraphics[width=0.7\textwidth,keepaspectratio,clip]{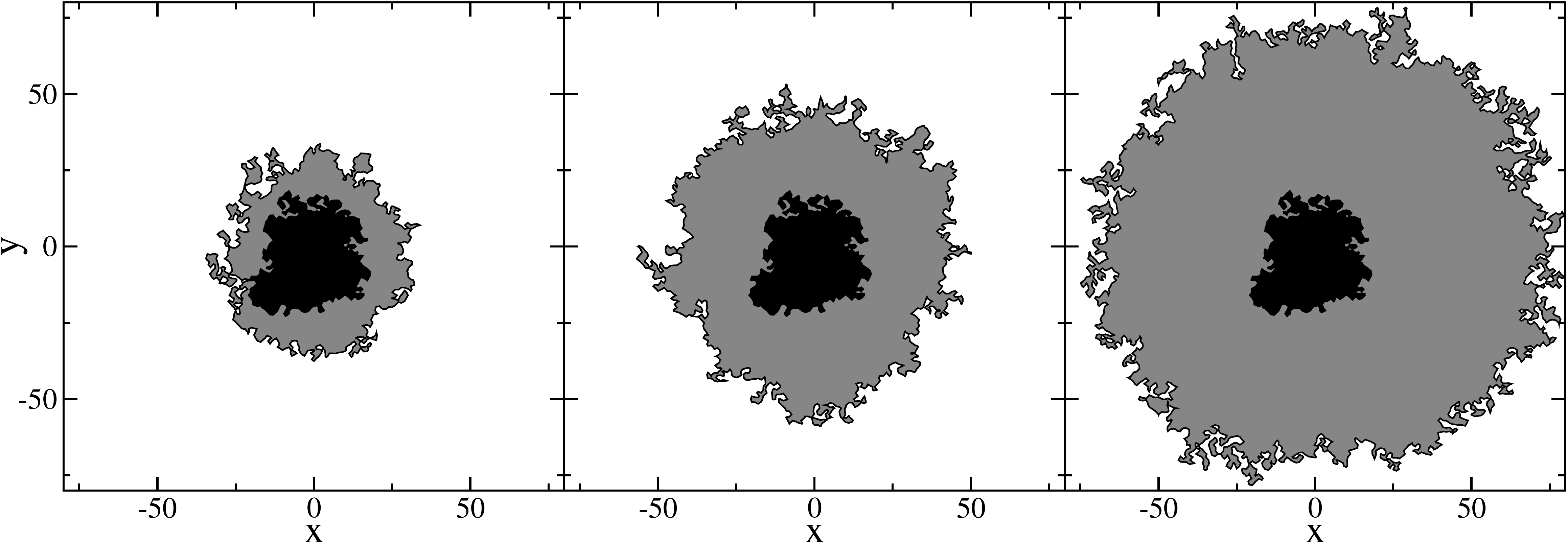}
\caption{Largest contour for $z_\rb \geq 10^{-5}$ at $t=10^6$. The black area in the center is the contour for the linear case, while the panels respectively correspond to (from left to right): weak chaos with $(\sigma,\mathcal{E})=(1.5,0.04)$, weak chaos with $(\sigma,\mathcal{E})=(0.5,0.00001)$, and strong chaos $(\sigma,\mathcal{E})=(0.5,0.005)$.} 
\label{fig:waves}
\end{figure*}

As a first response into this, we consider the normalized densities $z_\rb = \mathcal{E}_\rb / \mathcal{H}_K$ at  $t=10^6$, where an asymptotic regime is reached (cf. Figs.\ref{fig:sig_eq2}-\ref{fig:sig_le1}). In Fig.\ref{fig:waves}, the largest contour path for $z_\rb \geq 10^{-5}$ is shown, corresponding roughly to the expanding packet's surface. The left panel is for weak chaos with $\sigma>1$, the middle panel is for weak chaos with $\sigma\leq 1$, and the right panel is for strong chaos. Overlaid in black is the localized linear packet surface. A comparison of the coefficient $\alpha$ in Figs.\ref{fig:sig_gt1_lt2}-\ref{fig:sig_le1} can be observed. Wave packet in the left panel ($\alpha$ corresponds to weak chaos asymptotic) spreads less than presented in the middle panel ($\alpha$ in between the two limits), which spreads less than the right panel ($\alpha$ corresponds to strong chaos limit). However, the boundary shape itself provides no further evidence: no one boundary appears to be more fragmented or fingered than the others. 
\begin{table}[h]
\begin{center}
\begin{tabular}{|c|c|c|}
\hline
\rowcolor{Gray} Regime & $(\sigma, \mathcal{E})$ & $\avg{D_f}$ \\
\hline
Linear & 	N/A 	& 	$1.498\pm 0.045$ \\ \hline
\multirow{3}{*}{Weak Chaos, $\sigma>1$} & $(2.0,0.3)$ & $1.621\pm 0.019$ \\
				      & $(1.5,0.04)$ & $1.601\pm 0.021$ \\
				      & $(1.3,0.025)$ & $1.623\pm 0.017$ \\
\hline
\multirow{3}{*}{Weak Chaos, $\sigma \leq 1$} & $(1.0,0.006)$ & $1.667\pm 0.015$ \\
					     & $(0.7,0.0005)$ & $1.723\pm 0.011$ \\
					     & $(0.5,0.00001)$ & $1.740\pm 0.007$ \\
\hline
\multirow{2}{*}{Strong Chaos}		     & $(0.7, 0.030)$ & $1.734\pm 0.009$ \\
					     & $(0.5, 0.005)$ & $1.730\pm 0.007$ \\
\hline
\end{tabular}
\caption{Box-Counting Dimension for the different regimes, $z_\rb \geq 10^{-5}$.}
\label{tbl:BoxCntDim}
\end{center}
\end{table}

Therefore, we turn to a dimensional analysis of the contours. Binarizing $z_\rb$ at the the threshold $\geq 10^{-5}$, a box-counting algorithm \cite{moisy_computing_2008} is performed to extract the Minkowski-Bouligand dimension $D_f$ of the surface. The results were averaged over $100$ different realizations and presented in Table \ref{tbl:BoxCntDim}. This further suggests there are no clear and distinct fragmentation/fingering structures that might separate (in a geometric sense) the weak chaos at $\sigma \leq 1$ from the other two regimes.
\section{Conclusion}
We have investigated the spreading of a single-site excitation under a variable power nonlinearity within a 2-d disordered lattice, in particular for the Klein-Gordon case of Eq.(\ref{eq:HKG}). For such a system, we numerically confirm the second moment behavior of Eq.(\ref{eq:powers}), as first hypothesized in \cite{flach_spreading_2010}. In particular, we verify existence of both a weak chaos and a strong chaos regime. In addition, an intermediate regime for $\sigma \leq 1$ is observed, with spreading behavior between the two limits of strong and weak chaos. We have performed an analysis of the wavepacket geometries, but so far, strong fragmentation/fingering evidence in this regime, compared to the other regimes, is eluding. Possible future avenues along these lines may include lucunarity and density-density correlation measures. The behavior between the two regimes certainly remains open for future exploration, as well as pushing numerically the DNLS to achieve similar observations. 


\end{document}